%% file: manuscript.tex
\documentclass[a4paper,fleqn]{cas-dc}

\usepackage[numbers]{natbib}
\usepackage{bm}
\usepackage{xcolor}

\def\tsc#1{\csdef{#1}{\textsc{\lowercase{#1}}\xspace}}
\tsc{WGM}
\tsc{QE}
\tsc{EP}
\tsc{PMS}
\tsc{BEC}
\tsc{DE}
\begin{document}
\let\WriteBookmarks\relax
\def\floatpagepagefraction{1}
\def\textpagefraction{.001}

\begin{figure}[t!]
    \centering
    \includegraphics[width=0.9\textwidth]{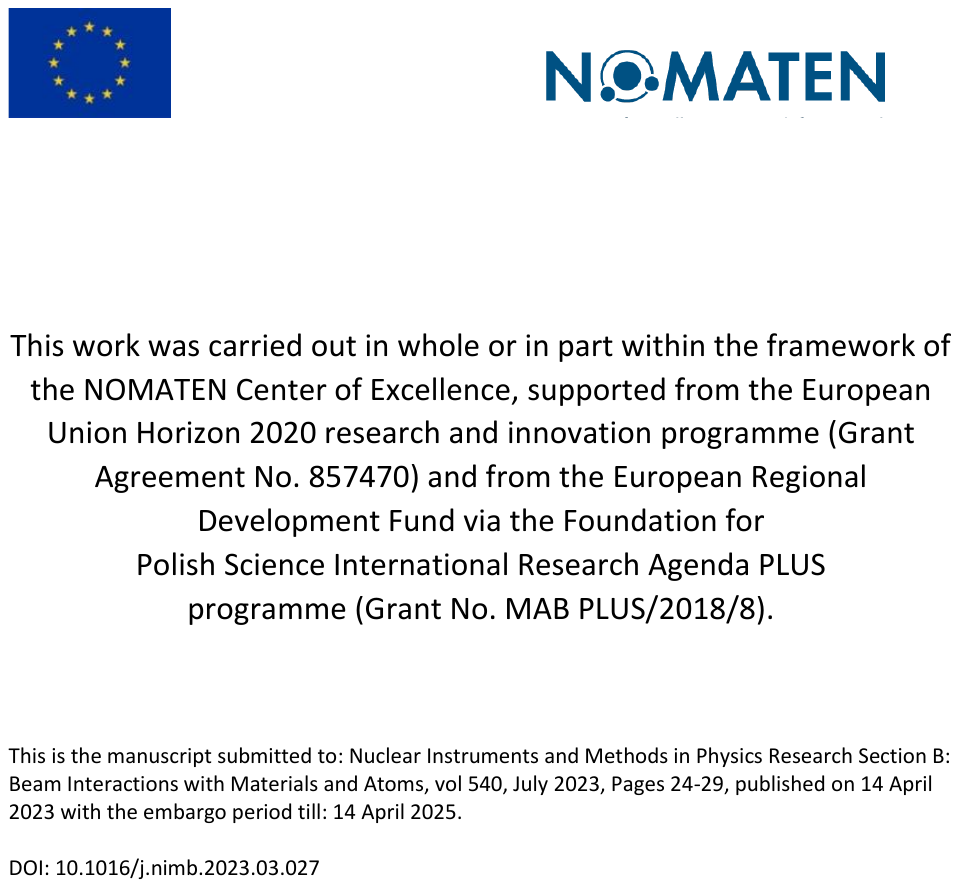}
\end{figure}

\shorttitle{Ion irradiation effects on hardening mechanisms of 
crystalline iron}

\shortauthors{K. Mulewska et~al.}

\title [mode = title]{ Self-ion irradiation effects 
on nanoindentation-induced plasticity of crystalline iron: A joint experimental and computational study}                      
%\tnotemark[1,2]

\author[1]{K. Mulewska}%[type=editor,
                        %auid=000,bioid=1,
                        %prefix=Sir,
                        %role=Researcher,
                        %orcid=0000-0001-7511-2910]
\cormark[1]
\credit{Writing Original draft preparation, \\ 
Conceptualization of this study, Methodology, Investigation, Visualization, Data curation}
\ead{katarzyna.mulewska@ncbj.gov.pl}

\author[1]{F. Rovaris}[orcid=0000-0002-0729-5409]
\credit{Writing-Original draft preparation, Methodology, Software, Visualization, Data curation}

\author[1]{F. J. Dominguez-Gutierrez}
\cormark[2]
\credit{Writing-Original draft preparation, Methodology, Software,Visualization, Data curation}
\ead{javier.dominguez@ncbj.gov.pl}

\author[1,2]{W. Y. Huo}
\credit{Methodology, Writing - Review \& Editing, Formal Analysis, Data curation}

\author[1]{D. Kalita}
\credit{Writing - Review \& Editing, Data curation}
\author[1]{I. Jozwik}
\credit{Writing - Review \& Editing, Formal analysis}

\author[1]{S. Papanikolaou}
\credit{Writing - Review \& Editing, Data Curation, Formal analysis}
\author[1,4]{M. J. Alava}
\credit{Writing - Review \& Editing, Data Curation}
\author[1]{L. Kurpaska}
\cormark[2]
\credit{Writing - Review \& Editing, Conceptualization of this study, Investigation, Formal analysis}
\ead{lukasz.kurpaska@ncbj.gov.pl}
\author [3,5]{J. Jagielski}
\credit{Writing - Review \& Editing, Data Curation}

\affiliation[1]{organization={NOMATEN Centre of Excellence, National Centre for 
Nuclear Research},
    addressline={ul. A. Soltana 7}, 
    city={Otwock},
    % citysep={}, % Uncomment if no comma needed between city and postcode
    postcode={05-400}, 
    country={Poland}}

\affiliation[2]{organization={College of Mechanical and Electrical Engineering, Nanjing Forestry University},
    city={Nanjing},
    % citysep={}, % Uncomment if no comma needed between city and postcode
    postcode={210037}, 
    %state={Trivandrum},
    country={China}}
\affiliation[3]{organization={Lukasiewicz Research Network
Institute of Microelectronics and Photonics},
    addressline={Wolczynska 133},
    city={Warsaw},
    % citysep={}, % Uncomment if no comma needed between city and postcode
      postcode={01-919}, 
    %state={Trivandrum},
    country={Poland}}
\affiliation[4]{organization={Department of Applied Physics, Aalto University},
    addressline={P.O. Box 11000}, 
    city={Aalto},
    % citysep={}, % Uncomment if no comma needed between city and postcode
    postcode={00076}, 
    %state={Trivandrum},
    country={Finland}}
\affiliation[5]{organization={National Centre for Nuclear Research},
    addressline={ul. A. Soltana 7}, 
    city={Otwock},
    % citysep={}, % Uncomment if no comma needed between city and postcode
    postcode={05-400}, 
    country={Poland}}
\cortext[cor1]{Principal corresponding author}
\cortext[cor2]{Corresponding author}

\begin{abstract}
In this paper, experimental work is supported by multi-scale 
numerical modeling to investigate
nanomechanical response of pristine and ion irradiated 
with Fe$^{2+}$ 
ions with energy 5 MeV high purity 
iron specimens by nanoindentation
and Electron Backscatter Diffraction. 
The appearance of a sudden displacement burst that is observed
during the loading process in the load--displacement curves is 
connected with increased shear stress
in a small subsurface volume due to dislocation slip activation and mobilization of pre-existing
dislocations by irradiation. 
The molecular dynamics (MD) and 3D-discrete dislocation dynamics
(3D–DDD) simulations are applied to model geometrically necessary dislocations (GNDs) nucleation
mechanisms at early stages of nanoindentation test; providing an insight to the mechanical response of
the material and its plastic instability and are in a qualitative agreement with GNDs density mapping images.
Finally, we noted that dislocations and defects nucleated  are responsible the
material hardness increase, as observed in recorded load–displacement 
curves and pop-ins analysis.

%Finally, described events allow us to understand better the behavior of the
%model material submitted to ion irradiation for applications in extreme 
%operating conditions.  
\end{abstract}

% \includegraphics{figs/grabs.pdf}

%\item Research highlights item 1
%\item Research highlights item 2
%\item Research highlights item 3

% Keywords

\maketitle

\input{sections/introduction}
\input{sections/exp_methods}
\input{sections/modelling_methods}

\input{sections/exp_results}

\input{sections/conclusions}

\section*{Acknowledgments}
%\ack
%We acknowledge 

This work was supported by the Euratom research and training
programme 2014-2018 under grant agreement no. 755039 (M4F
project) and has been supported by the EURATOM Direct Actions.
The research leading to these results was carried out in the frame of the Joint Programme on Nuclear Materials (JPNM) within the European Energy Research Alliance (EERA). 
This work is supported by the Ministry of Science and Higher Education through the Grant No 3908/H2020-Euratom/2018/2. 
We acknowledge support from the European Union Horizon 2020 research and innovation program under grant agreement no. 857470 and from the 
 European Regional Development Fund via the Foundation for Polish 
 Science International Research Agenda PLUS program grant 
 No. MAB PLUS/2018/8.
 We acknowledge the computational resources 
 provided by the High Performance Cluster at the National Centre 
 for Nuclear Research in Poland.
 The ion irradiations were carried out at the Ion Beam Center at Helmholtz-Zentrum Dresden-Rossendorf (HZDR).

\printcredits
\bibliographystyle{elsarticle-num}
\bibliography{references,references1}

\end{document}

%% file: sections/introduction.tex
\section{Introduction}
\label{sec:intro}
In recent years, with the development of new experimental techniques such as nanoindentation \cite {SCHUH200632, Minor, PhysRevB.73.054102} and pillar
compression\cite{HOSEMANN2012136, KIENER20062801}, tens and hundreds of nanometers have
become possible to be experimentally tested and analyzed.
At the same time, one can observe increasing computational
capacity and power, which allows for large scale simulation
by using molecular dynamics (MD) \cite{SZLUFARSKA200642,Javier2021, KURPASKA2022110639, Voyiadjis2017, Yaghoobi2014, Sato2020, Lee2020,VARILLAS2017431}
or discrete dislocation dynamics (DDD) \cite{JamondIJP2016, RovarisMSEA2022} of larger
material volumes. Combining these two routes enables us
to assume that experimentally observed crystal plasticity
instability can be visualized by in-situ or post-morten transmission electron microscopy (TEM) \cite{XU2020151911, STANGEBYE2022117441, GAGEL2016399}  following
dynamically the interaction of dislocations with microstructural features of the samples numerically. This offers a valuable opportunity to better understand the phenomena occurring in the tested material, their interaction, and the impact
of microstructural features such as grain boundaries, grain
orientation or precipitates on the mechanical properties such
as hardness or yield strength \cite{SCHUH200632, KURPASKA2022110639, PATHAK20151, MALERBA2021101051, RUIZMORENO2018168}. Along this
research line and goal, the combination of experimental and
numerical tools are essential for understanding the impact of
radiation damage and induced defect production \cite{NordlundNC}.

The ferritic/martensitic (F/M) steels are planned to be used in Generation IV nuclear reactors. They exhibit much lower swelling than austenitic steels, thus they can be utilized even up to 200 dpa \cite{HENRY2017329}. Those materials also display desired properties: higher thermal conductivity, lower thermal expansion, and a lower tendency to He embrittlement. However, F/M steels typically display significant hardening when exposed to radiation at temperatures around 300°C, the operating temperature of most reactors. In addition,  temperature fluctuations shall be unavoidable due to transients and temperature gradients (depending on reactor component particulars).  Hardening is commonly caused by creation of defects that pin and block dislocation movement, typically resulting in an increase of hardness and a decrease in fracture toughness, that is caused by localization of plastic deformation, with subsequent dramatic loss of uniform elongation.

To experimentally simulate radiation defects, one can
either perform neutron irradiation in the reactor core or use
an ion accelerator facility and perform ion irradiation. The second option is very tempting as an ion irradiation provides
an opportunity to produce controlled amounts of displacement damage under well-defined experimental conditions.
It is well established that ion irradiation can reproduce
all standard microstructural features observed in neutron--irradiated materials (dislocation loops, voids, and bubbles,
radiation-induced solute segregation, or radiation-induced
precipitates) \cite{PATHAK20151, MALERBA2021101051,Altstadt}. Charged particle irradiation in recent years has become valuable complementary tool to bulk
neutron irradiation studies \cite{KIM20095245, PhysRevLett.88.255507}. 
However, the possible
length scale of the irradiation damage region usually does
not exceed few $\mu$m for few MeV energies  \cite{PhysRevLett.100.135503}. 
Despite
this limitation, the development of experimental tools like
high resolution scanning electron microscopy/focused ion
beam (HR SEM/FIB) and TEM in the case of structural
investigations or nanoindentation allows for mechanically
testing sufficiently small volumes, which are relevant for ion irradiation conditions \cite{MALERBA2021101051}. Finally, one must remember that
the ion--irradiation offers opportunities for improved experimental control
that is impossible to perform during reactor
neutron irradiation such as separate effects testing. Therefore, the separation of different environmental conditions
is much easier, hence understanding individual phenomena
becomes possible One of these phenomena is the impact of radiation defects on the dislocation nucleation process and evolution. 

It
has been commonly agreed that the nanoindentation method
is one of the key techniques in understanding plasticity in
small volumes \cite{SCHUH200632, PhysRevB.73.054102,RUIZMORENO2018168,Linature}. A common agreement has
been achieved that this technique can be used to study fundamental dislocation nucleation aspects. The advantage of
this technique is the ability to probe small volumes, limiting effect of different types of microstructural defects. Suppose
one is testing single crystals or individual, large grains. In
that case, it can be assumed that the onset of plasticity
is marked by the characteristic discontinuity in the 
load-indentation depth (L-D) curves, 
called a “pop--in” event. These
events are generally explained by homogeneous dislocation
nucleation. It is also assumed that these effects occur when
shear stress generated under the indenter tip approaches the
material’s theoretical shear strength. When the first pop-in
effect is observed [18], which means that the homogeneous
dislocation nucleation in the defect-free (this is valid for
single crystals or grains probed sufficiently far from the
grain boundary) volume occurred \cite{MALERBA2021101051}. 
The evolution of
the dislocation can be better understood if an computational
model is applied at different length-time scales. This has
been well documented for face centered cubic (FCC) systems. 
However, dislocation motion and interaction in body
centered cubic (BCC) are still poorly understood. Several
features such as interstitial atoms, surface roughness, crystal
orientation, or existing dislocation density and their interaction are to be explained \cite{VARILLAS2017431}. 
The situation is even more
complicated for the irradiated system when radiation defects
are introduced to the material, and they evolve in time and
scale.

Considering all the above aspects, the present paper
aims to better understand the effect of incipient plasticity
in BCC-type high purity Fe (model material). Our experimental 
and numerical studies are performed on pristine and
ion-irradiated materials. We consider one-grain orientation
and analyze the possible correlation between the crystallographic orientation, presence of radiation defects, and
occurrence of pop-in events. We extrapolate our atomistic
simulation based on molecular dynamics (MD) method by
performing discrete dislocation dynamics (DDD) simulations to investigate the evolution of dislocation forest at
larger material volumes during nanoindentation testing.
DDD and MD simulations are both useful computational
methods that can be used together to gain deeper understanding
of materials behavior. DDD simulations focus on mesoscopic features of materials by modeling the movement
and interactions of dislocations, where nucleation and multiplication of dislocations 
is captured on the level of dislocations, but not on the atomistic scale.  New dislocation segments are naturally introduced by 
multiplication of pre-existing dislocations. 
Atomistic MD simulations can guide DDD approaches by 
describing initial nucleation of dislocations at 
the onset of plasticity. 
Herein, the microstructure and mechanical properties of
the Fe sample are studied. The results are supported by
numerical modeling. The dislocation nucleation mechanism
is proposed via a computational model. It is compared to
high resolution electron backscatter diffraction (HR-EBSD)
orientation image analysis, and shows information about the
nanomechanical response of the Fe sample under external
loads.

%% file: sections/exp_methods.tex
\section{Experimental and computational methods}
\label{sec:methods}

\begin{table*}[width=1.99\linewidth,cols=10,pos=b]
\caption{Chemical composition of iron samples.}\label{tab:Tab1}
\begin{tabular*}{\tblwidth}{@{} L|LLLLLLLLL@{} }
\toprule
Element   & C & Cr & Ni  & Si
& P & Al& Fe\\
\midrule
Weight \% & <0.005 & 0.002 & 0.007 & 0.001 
& 0.003 & 0.013 & Bal.\\
\bottomrule
\end{tabular*}
\end{table*}

%%%%%%%%%%%%%%%%%%%%%%%%%%%%%%%%%%%%%%%%%%%%
%\subsection{Experimental methods}
%============================================================
\subsection{Experimental nanoindentation technique}
Nanomechanical investigations were carried out by using 
a NanoTest Vantage system provided by Micro Materials Ltd.
Measurements were performed at room temperature with a 
Berkovich-shaped diamond indenter tip. 
It is known that tip place a major role during low load 
nanoindentation \cite{FISCHERCRIPPS20064153}.
Prior to the tests, the diamond area function (DAF) of the 
indenter was calibrated using fused silica with 
known mechanical properties.
Indentations were performed in single force mode with
a maximum load of 1.5 mN, 
which corresponds to 200 nm on pristine and 
130 nm on irradiated samples. According to generally accepted 
nanoindentation principles, the plastic deformation regime 
induced during indentation, is 5--10 times 
greater than the penetration depth, depending on the material
being tested \cite{CHEN2009911}. 
This effect is less drastic in harder materials, with high 
yield strength. 
However in our studies, we adopted the most restrictive 
criterion to ensure that the material response comes from 
the irradiated layer, whose peak damage is located at a 
depth of around 1.3~$\mu\text{m}$.
The distance between indents was set at 10 $\mu$m. 
For clarity, occasional faulty trials were manually excluded.  

%========================================================
\subsection{Electron backscatter diffraction (EBSD) analysis}

A HR--EBSD analysis, in terms of local orientation gradients in
grains deformed by nanoindentation, was performed  using a Helios
5 UX (ThermoFisher Scientific) field emission scanning electron
microscope operated at 25 kV and 6.4 nA current. 
A Kikuchi diffraction pattern was collected for each point at
a camera resolution of $640 \times 480$ pixels and a step size
of 30 nm using EDAX Velocity Pro EBSD system. 
The orientation maps were analysed using OIM Analysis software
with the application of a confidence index (CI) standardization
routine and the filtering of the data with CI below $0.1$.

%%%% Material
\subsection{Preparation of the pristine and irradiated material}
\begin{figure}[!b]
   \centering
   \includegraphics[width=0.4\textwidth]{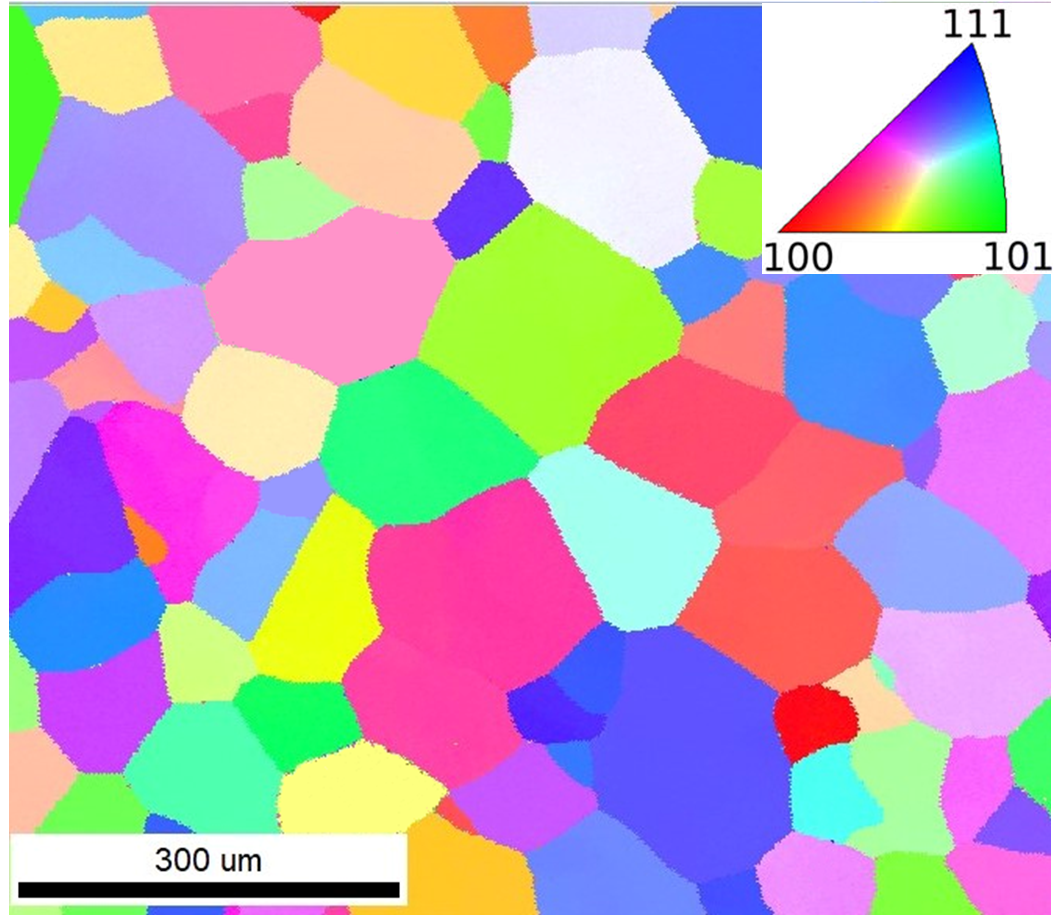}
   \caption{(Color online) EBSD IPF Z image of the pristine Fe sample.}
   \label{fig:EBSD}
\end{figure}

The studied model material of high purity iron used in
this work was cast by OCAS (Gent, Belgium) in an 
induction vacuum furnace by additive melting. 
The chemical composition is reported in Tab. \ref{tab:Tab1}.
%, not listed elements were below the detection limit. 
The studied model material was fabricated as a rolled
plate with a final thickness of 10 mm. 
This was done by cutting a $50$ mm $\times$ $125$ mm 
$\times$ $250$ mm piece from each lab cast. 
Then these pieces were introduced to the pre--heated 
furnace at $1200^o$C for 90 min. 
Following this step, hot rolling was performed 
(in 7 passes) until $10$ mm was achieved. 
The hot rolling lasted for about $60$ s, and the 
temperature of each element was approx. $930^o$C 
after the termination of the process. 
Afterward, cooling to room temperature in air was
performed. 
The final dimensions of the sheets (measured in the
rolling direction) were about 10 mm $\times$ 250 mm $\times$ 600 mm 
(height $\times$ width $\times$ length).
The samples surface were mechanical polish at step--wise decreasing
grain size of the polishing agent. One can observe in the 
EBSD IPF Z image (Fig. \ref{fig:EBSD}) that the studied specimens have 
large globular in shape grains with an average size of approximately
90 $\mu$m with fully ferritic microstructure and randomly 
distributed crystal orientation.

%%%% Material

%%%%% Figure adding

\begin{figure}[!t]
   \centering
   \includegraphics[width=0.48\textwidth]{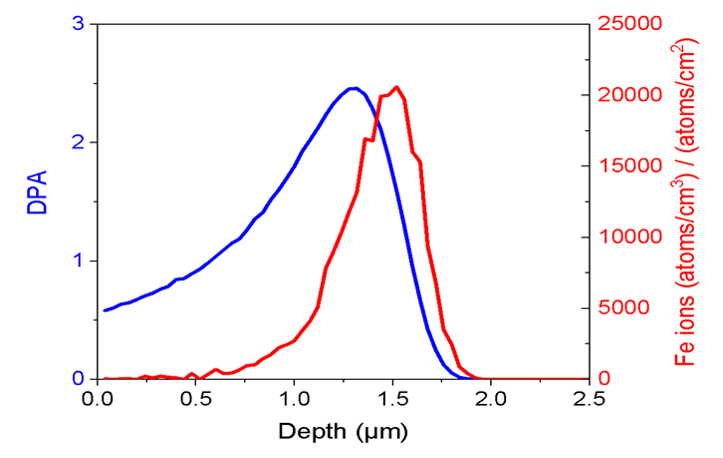}
   \caption{(Color online) Profiles of displacement damage (blue line) and concentration of injected interstitials (red line) for the 5 MeV $Fe^{2+}$-ion irradiations, obtained by SRIM calculations.}
   \label{fig:srim}
\end{figure}

Ion irradiation was performed with the 3 MV 
tandetron accelerator located at the Ion Beam Center
at HZDR Dresden. 
Fe$^{2+}$ ions with 5 MeV energy were implanted at 
300$^o$C into the sample. 
The temperature control was based on a thermo--couple
placed on the backside of the sample, while the
specimen was mounted on the heating target. 
The samples were scanned by a properly focused ion 
beam such that the irradiated area received a 
laterally uniform exposure corresponding to the 
respective predetermined target values. 
The ion flux was monitored continuously utilizing
Faraday cups and integrated to obtain the ion
fluence. 
The profiles of displacement damage in units of
displacements per atom (dpa) and injected 
interstitials per atom (ipa) were calculated 
using the SRIM binary collision code. 
This was done according to the recommendations 
given by Stoller et al.\cite{STOLLER201375} using the quick
Kinchin-Pease calculation and displacement
energy of 40 eV. 
The respective profiles are plotted in Fig. 
\ref{fig:srim}.

 %%%%% Figure adding

%%%%% Figure adding
%\begin{figure}[!t]
%   \centering
%   \includegraphics[width=0.48\textwidth]{Figures/EBSD_indenty.png}
%   \caption{EBSD map with indents}
%   \label{fig:fig3}
%\end{figure}

%% file: sections/modelling_methods.tex
\subsection{Computational modeling}

\subsubsection{Molecular dynamics simulations framework}

MD simulations is a used to describe the physical processes
during nanoindentation testing, at an atomic level. 
For this, the Large-scale
Atomic Molecular Massively Parallel Simulator (LAMMPS) 
\cite{THOMPSON2022108171} and 
interatomic potentials based on EAM method
\cite{EICH2015185} were performed to describe Fe-Fe interaction. 
The modelling starts by the preparation of a single crystal
$[101]$ Fe sample with 11664000 Fe atoms in a $(51.53,54.66,48.58)$ 
nm cell with density of 7.92 g/cm$^3$, which agrees well 
with the experimental data. The size of the sample 
was chosen to be large enough for the anticipated propagation of dislocations along slip planes and dislocation dynamics throughout the BCC sample during the 
indentation process. Naturally, this requirement results in larger dimensions along the $z$--direction.
Further, a process of energy optimization follows,
and equilibration for 100 ps with a Langevin thermostat
at room temperature.
In order to emulate nanoindentation, the Fe sample is prepared 
by following the process explained in our previous work 
\cite{Javier2021,KURPASKA2022110639,Dominguez-Gutierrez_2022}.
We utilize a repulsive imaginary (RI) rigid sphere 
as our indenter tip, with radius $R = 10~\text{nm}$ and 
a speed $v = 20~\text{m/s}$, where $v$ is chosen as positive
for loading, and negative for unloading processes.
Each calculation was performed for 125 ps with a time step
of $\Delta t = 0.5$ fs. for a maximum indentation depth of
$4.0~\text{nm}$ to avoid the influence of boundary layers of the
material. This methodology was used in our previous works 
to model nanoindentation, in good agreement with experimental
data
\cite{Javier2021,KURPASKA2022110639}
Thus, the geometrically necessary dislocations (GNDs) are computed 
by using DXA method \cite{ovito} implemented in OVITO~\cite{Stukowski_2012}.

%%%%%%%%%%%%%%%%%%%%%%%% 3D-DDD approach
\subsubsection{3D-DDD approach}

The DDD approach applied in this paper exploits a coupling between a Dislocation Dynamics code~\cite{DevincreBOOK2011} and 
a Finite Element (FE) solver~\cite{FenicsBOOK2016} for consistently evaluating the dislocation stress field near the free surfaces. The coupling is based on the eigenstrain formalism~\cite{MuraBOOK1987}, and implemented following the 
Discrete-Continuous Model (DCM) scheme~\cite{VattreJMPS2014,JamondIJP2016,RovarisPRM2017,RovarisMSEA2022}.

In the DCM, the stress field acting on dislocation segments is calculated by solving the Partial Differential (PD) equation of mechanical equilibrium. In this approach the plastic deformation is introduced in the constitutive equation by means of the eigenstrain formalism and a properly chosen regularization function~\cite{CaiJMPS2006}. This allows for the coupling with a FE numerical solver. The equilibrium equation to be solved thus become:  
\begin{equation}
	\begin{cases} 
		-\nabla\vec{\sigma}(\vec{u}) =  \vec{0} 								\,\,\quad\qquad\qquad\qquad \text{on $\Omega\setminus\partial\Omega$}\\
		\vec{\sigma}(\vec{u}) = C(\vec{\epsilon}(\vec{u}) - \vec{\epsilon}^*) \,\qquad\qquad \text{on $\Omega\setminus\partial\Omega$}\\
		\vec{u} = \vec{0}				     						 		\qquad\qquad\qquad\qquad\qquad \text{on $\partial\Omega_{\text{D}}$}\\
		\vec{\sigma}\cdot \vec{n} = \vec{0} 	   						  		\,\,\,\,\qquad\qquad\qquad\qquad \text{on $\partial\Omega_{\text{N}}$}\\
		\vec{\sigma}\cdot \vec{n} - k_{\text{pen}}<u-Z>_{+}= \vec{0} 	   		\quad \text{on $\partial\Omega_{\text{c}}$}\\
	\end{cases}
	\label{eq::equil}
\end{equation}
where $\bm{u}$ is the unknown displacement field, $\epsilon^*$ the eigenstrain, representing the plastic deformation, $\Omega$ is the simulation volume and $\partial\Omega$ its external boundary. The bottom boundary $\partial\Omega_{\text{D}}$ is kept fixed by applying  Dirichlet boundary conditions (zero displacement), while  $\partial\Omega_{\text{N}}$ is the top free surface with Neumann boundary conditions (zero normal stress). The last condition in Eq.~\eqref{eq::equil}, is applied to the contact region $\partial\Omega_{\text{c}}$. The constant $k_{\text{pen}}$ is a penalization constant and the symbol $<>_+$ denotes the positive part of the argument. 
The function $Z$ is used to define the outer surface of the indenter, modeled as the parabolic approximation of a sphere in this work.

In this way, boundary conditions for free surfaces of arbitrary geometry are fully taken into account when solving the PD system of Eqs.~\eqref{eq::equil}. Finally, the stress field resulting from the FE solution is used to compute forces that move dislocations at the $N\text{\emph{th}}+1$ time step, following an iterative process. It is important to highlight that, while the FE solver is in charge of computing the mechanical equilibrium, the DD code not only handles the movement of dislocation segments but it also manages potential reactions between them, modeling local interactions and recombinations. When two dislocation segments fall inside the regularized region their interaction is evaluated by means of the analytical expression derived in Ref.~\cite{CaiJMPS2006}, consistently exploiting the same regularization function $\tilde{w}$. Thus, the dynamics described by the DCM model is independent on the choice of the regularization thickness~$\tilde{h}$~\cite{JamondIJP2016}.

For the purpose of this paper, simulation boxes of 1~$\mu$m lateral sizes with lateral periodic boundary conditions were considered. 
The bottom boundary of the simulation cell is kept fixed while the top one is a free surface, as defined in Eq.~\eqref{eq::equil}. A distribution of pre-existing prismatic dislocation loops were placed under the indenter location to act as a trigger for the plastic deformation during nanoindentation. These loops were randomly selected between all possible glide systems for BCC Fe and with random positions in a small volume box (around $100~\text{nm}$ of lateral size) under the desired indenter location. Multiple simulations have been performed in order to confirm that details of the dislocation network at late indentation stages did not depend on the details of the initial prismatic loops distribution.

%% file: sections/exp_results.tex
\section{Results and discussion}
\label{sec:results}
%%%%%%%%%%%%%%%%%%%%%%%%%%%%%%%
%%%%%%%%%%%%%%%%%%%%%%%%%%%%%%%

%%%%% Figure adding
\begin{figure}[!b]
   \centering
   \includegraphics[width=230pt,height=210pt]{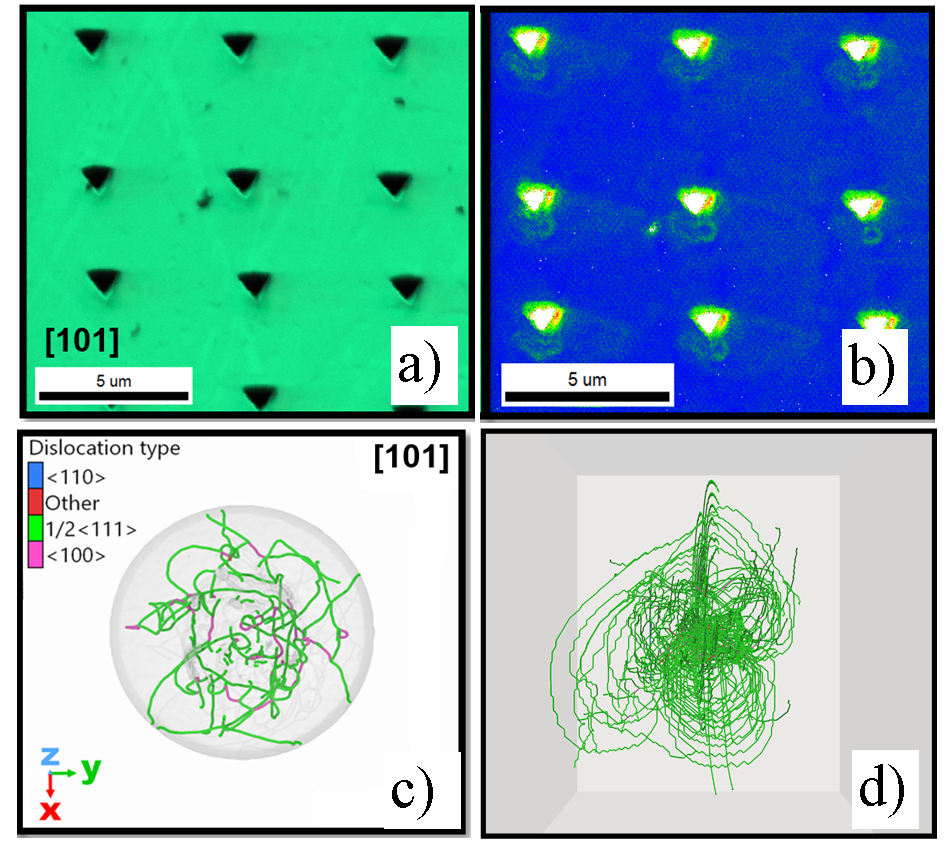}
   \caption{(Color online). a) EBSD map of the indentation 
   array made within one grain with [101] orientation, b) 
   GND density of map of the indentation array, and c) calculated dislocation network at the maximum 
   indentation depth calculated for the spherical
   indentation up to 4 nm depth; and d) 3D-DDD results 
   for dislocation network following the coordinate tripo and dislocation 
   color palette shown for the MD results.}
   \label{fig:fig4}
\end{figure}

The dislocation nucleation process
plays a dominant role in the yield and subsequent
plastic deformation of the material. 
To better understand the onset of plasticity by dislocation nucleation for the pristine case, the nanoindentation array was observed from the point of a specific grain. 
Figure \ref{fig:fig4}a) shows an array of indents 
made on a grain with a preferential orientation of
[101]. 
The test was performed on the pristine specimen. 
We analyzed the recorded HR-EBSD image and 
calculated the density of geometrically necessary
dislocations (GNDs) on the surface near the indent. 
More GNDs are visible near the indenter 
tip and on one side, on the direction \{110\} slip plane. 
To better understand the mechanism of the loop 
formation, MD was carried out to simulate the 
indentation process. 
Fig. \ref{fig:fig4}c) shows the developed dislocation 
types created in the $[101]$ grain at the maximum indentation 
depth provide information of the dislocation forest underneath 
the tip. 
As expected, $1/2 \langle111\rangle$ loops are most common. 
In addition, the presence of $\langle 110 \rangle$  and
$\langle 100 \rangle$ dislocation segments can be noticed. 
A very similar behavior has also been reproduced by means of
DDD simulations, as reported in Fig.~\ref{fig:fig4}d). 
Here,
the simulation box is much larger due to the computational
advantages of DDD simulations and the maximum indentation depth
simulated is much closer to the experimental one, 
around $55~\text{nm}$. 
The dislocation network represents the same features observed
in the MD simulation and in the experiments 
(Fig~\ref{fig:fig4}(b) and (c)) with the main network being
formed by $1/2 \langle111\rangle$ dislocation loops. 
Some dislocation-dislocation reactions occur mainly forming
$\langle 100 \rangle$ dislocation segments and maintain their evolution 
as reported by DDD simulations (Fig. \ref{fig:fig3}d).

%%%%% Figure adding
\begin{figure*}[!b]
   \centering
   \includegraphics[width=\textwidth]{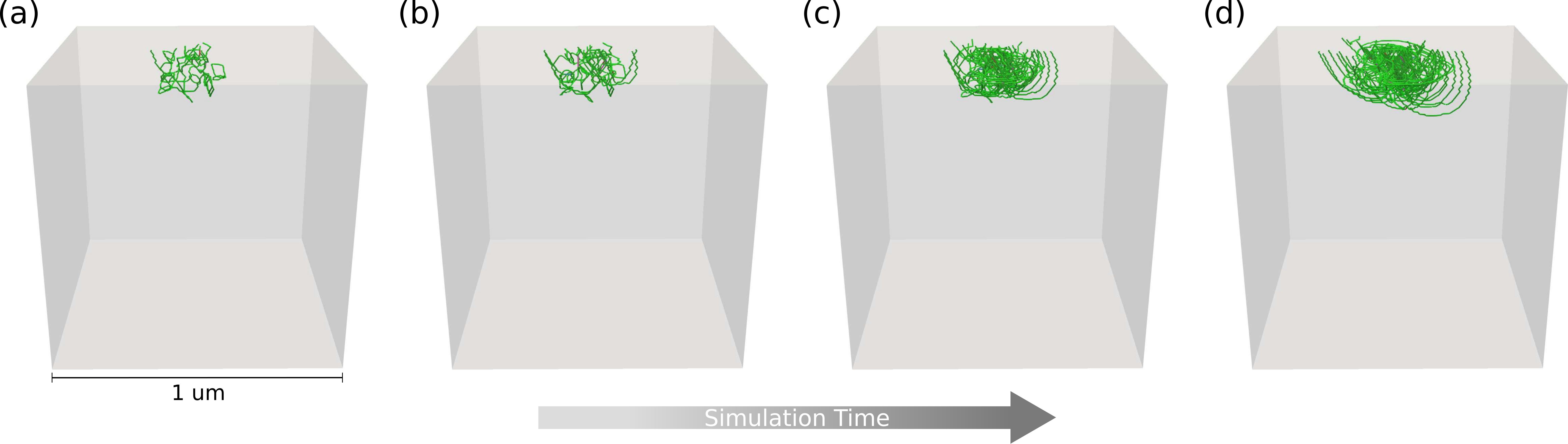}
   \caption{(Color online). Selected snapshots from a DDD 
   simulation of nanoindentation of pure Fe in $[101]$ 
   directions. 
   The initial configuration consists of randomly placed
   prismatic dislocation loops under the indenter as described
   in the text. 
   The simulation times are 2 ns (a), 4 ns (b) 6 ns (c) and 
   8 ns (d). The maximum indentation depth shown in (d) 
   corresponds to approximately $55~\text{nm}$.}
   \label{fig:fig5}
\end{figure*}
 
The temporal evolution of the dislocation microstructure can 
be better observed in Fig.~\ref{fig:fig5} where 
selected snapshots from the loading process are extracted 
from a DDD simulation. 
As discussed in Section~\ref{sec:methods}, the starting 
conditions of the DDD simulations performed here is a random 
distribution of prismatic loops placed under the indenter 
location. 
From this starting configuration the simulations evolve as
shown for examples in Fig~\ref{fig:fig5} (a)-(d). 
The dislocation loops rearrange under the influence of the 
indenter stress fields and start extending on their glide 
planes, multiplicating and reacting in order to form the 
dislocation microstructure already described in 
Fig.~\ref{fig:fig4}. 
The maximum indentation depth achieved in this simulation
is approximately $55~\text{nm}$ and the corresponding dislocation
microstructures have already been reported in the top view
of Fig.~\ref{fig:fig4} (d).

Figure \ref{fig:fig3}a) shows, recorded during indentation, load-indentation depth 
curves on pristine (black curves) and ion irradiated 
(blue curves) specimens. 
Indentations were performed with a constant load up to 
max 1.5 mN, resulting in deforming approx. 200 nm in
the case of pristine and 130 nm in the case of ion
irradiated material. This load was chosen to assure
the probing volume of the material, which does not
extend below the irradiated region (even assuming a
very constrained 1/10 criterion). 
Figure \ref{fig:fig3}a) presents all recorded 200 indentation curves,
hence, at first glance, the observed pop-in effect may
be hidden. 

Figure \ref{fig:fig3}b) shows representative loading parts 
recorded during the indentation test. 
One can identify the occurrence of two clearly 
isolated pop-ins (with different magnitudes) in 
both cases. As described in the introduction section,
the occurrence of a first pop--in can be attributed to
the transition from purely elastic behavior to the
onset of additional plastic deformation. 
All recorded curves show at least one pop--in event. 
This proves that the dislocation activation process
followed by dislocation nucleation. 
This is a classical behavior for BCC metallic
materials, and the recorded results are in full
agreement with the experimental work of Ahn et al. \cite{ahn_oh_lee_george_han_2012}
and numerical data provided by Biener et al. 
\cite{PhysRevB.76.165422}.

%%%%%%%%%%%%%%%%%%%%
%%%%% Figure adding
\begin{figure*}[!b]
   \centering
   \includegraphics[width=\textwidth]{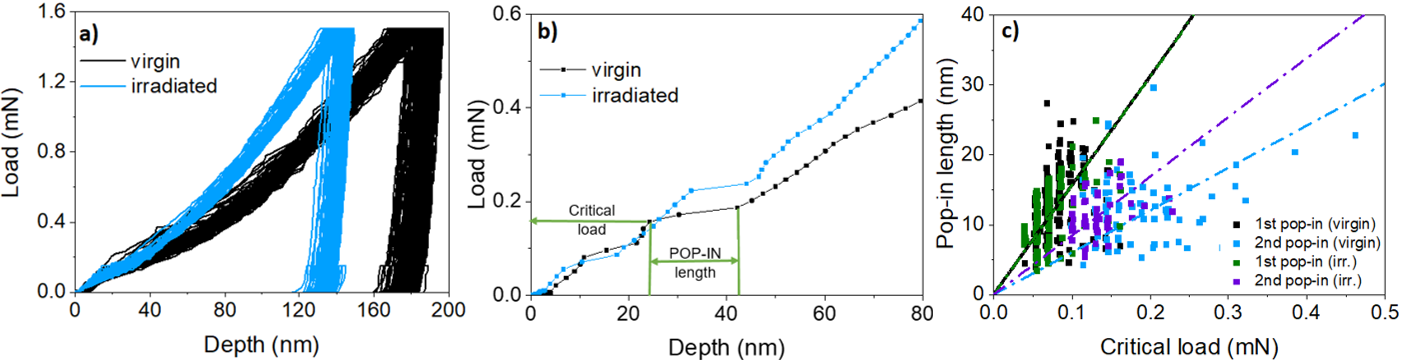}
   \caption{(Color online) a) Experimentally recorded L--D curves
   of pristine (black line) and ion irradiated (blue line)
   pure Fe-specimen. 
   Indentations were performed up to the maximal load of $1.5$ mN, which
   assured probing only radiation damage zone, 
   b) representative loading part of the L-D curves marked with arrows, 
   sudden displacement bursts represent the first and secondary pop-in 
   effect recorded for both samples, and c) correlation between pop-in 
   initiation load and length recorded for the first and secondary pop-in
   effect.}
   \label{fig:fig3}
\end{figure*}
 
To quantify the occurring pop-in event, we calculated 
the number of pop-ins, length of the displacement, and
critical load upon which this event commenced. 
This was done for both pristine and ion irradiated
samples. Regardless of the state of the sample, a
similar trend between the 1st and 2nd pop-ins has been
observed. 
The first pop-in always occurs at a lower critical
load and is statistically more significant than the
second event. 
When considering only the first pop-in event, one can
see that no difference between pristine and
ion-irradiated material exists. 
In both cases, it requires approximately 10 $\mu$N load 
to activate the event \cite{MALERBA2021101051}. 
This proves that both samples have the same 
dislocation density and number of pre-existing 
mobile dislocations in the shear zone and have 
the same residual stress level at the surface 
after polishing. 
As demonstrated by P\"ohl \cite{Poehl}, polishing to high 
roughness level results in the development of 
isolated small pop-ins. 
This proved that local stress concentrations
caused by surface discontinuity favor nucleation
or activation of the pre-existing dislocations, 
which leads to the disappearance of the pop-in. 
We did not observe such phenomena in our case. 
Therefore, one can conclude that the roughness 
of the samples (and stress) are at the same level.
However, one can see that a significant difference
exists when comparing second pop-ins. 
Blue points represent the magnitude of the pop-in
in the pristine sample, while purple points show
the behavior of the ion irradiated sample. 
Statistically speaking, it is clearly seen that
ion-irradiated samples shows a more consistent
response. 
In addition, one can see that the first
and second pop-ins recorded for pristine samples
(see black and blue points on Fig. \ref{fig:fig3}c) 
are slightly
larger both in terms of the critical load necessary
to active the event and in pop-in length. 
It is known that in pure iron and BCC alloys
generally, the mobility of pre-existing dislocations
can be influenced by interstitial atoms such
as carbon. 
One must remember that C is often introduced into
the system during irradiation. In addition, the ion 
irradiation campaign was performed at $300^{\circ}\text{C}$, which
is sufficient for C atoms to migrate and redistribute.
Dissolved carbon atoms may diffuse into the stress fields of 
pre-existing dislocation, where they form the so-called 
Cottrel atmosphere, which results in dislocation
pinning \cite{SEKIDO2011396}. 
This impedes their mobility and can be observed as 
a much smaller amount of pop-ins and their lesser 
magnitude in ion irradiated material. 
Therefore, we believe that a significant reduction
in pop-in magnitude is related partially to this
effect. This is consistent with the work of
Barnoush\cite{BARNOUSH20121268}. 
The second phenomenon responsible for
lowering the magnitude of the pop-in is the 
presence of dislocation loops generated due to 
the development of radiation defects.

%% file: sections/conclusions.tex
\section{Concluding remarks}
\label{sec:conclusion}

In this work, we carried out a joint experimental, and multi--scale 
computational study of plastic deformation mechanisms of pristine 
and ion irradiated polycrystalline BCC iron through 
nanoindentation. 
For the pristine case, we characterized and described defect 
nucleation during nanomechanical testing and the evolution of 
geometrically necessary dislocations densities (GNDs) in
[101] grain indentation. To better understand the evolution during dislocation 
nucleation, we performed MD and 3D-DDD simulations, that showed the formation of GNDs, 
mainly with a Burgers vectors  b= 1/2⟨111⟩, in agreement with experiments.
Furthermore, we investigated the pop-in behavior during nanoindentation.
The analysis of L--D curves showed the occurrence of two sequential pop--ins 
in both cases studied (pristine, irradiated). The first 
pop--in is observed on both virgin and ion irradiated
surfaces. 
In the case of the second pop-in, however, the irradiated sample required significantly higher stress for its activation, a phenomenon that may suggest that carbon was
introduced into the material during ion irradiation at 300°C. At this temperature, carbon atoms may diffuse and form Cottrell atmospheres, 
resulting in dislocation pinning.